# Transformations of mass/charges fission fragments spectra with consideration the post-fission emission of nuclear particles: $^{232}$Th

V. T. Maslyuk, O. O. Parlag, M. I. Romanyuk, O. I. Lendyel and O. M. Pop

Institute of Electron Physics - Universitetska str., 21, 88017 Uzhhorod, Ukraine

The possibilities of the proposed statistical approach are shown in the task of investigating the post-scission transformation of mass and charge yields of fission fragments with considering the emission of different length of chains of elementary particles as the beta particles ($^{\pm}\beta$) and neutrons fission. Using as the example the $^{232}$Th isotope, which is located on the border between pre-actinide and actinides nuclei, the changes in the topology of fission fragments yields, the role of emission of nuclear particles in the competition of symmetric and asymmetric modes in fission processes are investigated. The proposed method allows one to determine the most probable two fission's fragments clusters, obtained in considering the different chain's length of emitted nuclear particles and the parameters of their stability. All these results are essential when interpreting the experimental data.

## 1. Introduction

The mass/charge spectra of fission fragments (MCSFF) yields are important observable parameters that are dependent on the peculiarities on the fission process of atomic nuclei both as a spontaneous or stimulated ones, on the nature of the internuclear interaction, and on the influences of filled nuclear (proton, neutron) shells [1]. The study of the MCSFF provides valuable information on the stability of nuclear matter, such tasks follow from the physics of nucleosynthesis, the interaction of beams of radioactive high-energy ions with the possibility of synthesis of new super-heavy and unstable nuclei [2-4]. Fission fragments are the basis for the formation of the chemical and microelement composition of the earth's crust and space objects, and their study is essential for radioecology tasks when forming the parameters of "radiation weather" [5], etc. The applied significance of studying the fission fragments yields is due to many applications, from the needs to control the quality of nuclear fuel, the storage of radioactive waste to the tasks of nuclear medicine and environmental studies.

However, in practice, it is difficult to establish the MCSFF because a post - scission ensemble is a dynamic system and influenced by the different chains of transformation due to the emission of neutrons (prompt, delayed) or beta particles [6]. The inverse problem of changing the ensemble of nuclear fragments may occur in r-nucleosynthesis processes when the fission fragments can absorb neutrons in the amalgamation of binary neutron stars that can also change their composition [7]. When we take into account that the time scale of such processes includes ultra-short-lived and almost stable nuclei-fragments, it becomes essential to estimate the reliabilities of the obtained MCSFF within the abilities of the widely used experimental methods as radiochemical, mass- and gamma-spectrometric. Investigation the nature of the time evolution of the MCSFF' ensemble, for example, when one considers the post-scission emission of elementary particles, is also a difficult task for theoretical studies. Now, there are exist of different theoretical approaches to describe the formation of fission fragments yield and the topology of their mass-charge spectra based as a rule on various modifications of the liquid drop model. These theories are using the idea of restricted Fermi gas\fluid and widely exploit not only the formal mechanical as size, shape, viscosity but also the kinetics characteristics of the atomic nuclei's scission processes [8-10]. However within this approaches the post-scission state of the nucleus is considered as the binuclear system only, and the process of their nuclear transformations are neglected. The progress achieved in such researches is realized in the numerical codes as GIS and others [9]. As a rule, such approaches take into account the stage of inelastic nuclear interaction, the processes of thermalization and statistical relaxation of the initial nucleus, and, finally, the formation of its decay channels without emission of elementary particles. Currently, there are tendencies to go beyond the simple two-fragments scission scheme and to take into account the presence of cluster radioactivity, for example, due to the emission of alpha particles [11], as well as the effects caused by the emission of nuclear particles from the fission fragments. Such theories would allow to investigate the time dependence of the MCSFF transformations and to explain the experimental data of the nucleus fission' studies when the systematization of fission fragments is carried out only after some time after the nucleus fission and the evolution of the ensemble of the fragments of their nuclei.

In this paper, we for the first time propose a method for studying the transformation of MCSFF taking into account the emission of nuclear particles, in particular, the different length's chains of neutrons of fission and beta particles within the framework of a new statistical approach that allows one to study the ordering of the post-fission ensemble of fission fragment clusters. According to [12], the state of such ensemble of nucleons can be described by thermodynamic parameters: temperature, T and pressure P, whose values are determined by the characteristics of the initial or fissile nucleus. Such an approach does not require the introduction of fitting parameters and is universal in studying of the post fission characteristics of both light (pre-actinide) and super-heavy nuclei. A similar ideology is implemented also in the SPY (Scission Point Yields) model, which focuses on [13] studying the yields and energy characteristics of fission fragments, based on the analysis of all statistical nucleon configurations that can be realized at the nuclear fission point.

The calculation was carried out using the example of the fission fragments of the $^{232}$Th isotope, which is on the verge of pre-actinides and actinides nuclei and is of interest for practical applications in nuclear energy of the future.

## 2. Theory

The basics of the post-scission approaching and the proposed statistical method for the systematization of fission fragments are described earlier in [12, 14-16]. In its framework, the nuclei are considered as a condensed state of nuclear matter, for which the binding energy is not an additive function according to the sort (protons/neutrons) and a number of nucleons.

The subject of the study is the initial nucleus with atomic mass $A_0$, and the charge $Z_0$, and its fission fragments that had been formed after the passing the saddle point, their mass/charge ratio, and proton/neutron content. In such assumption, fission fragments and their nucleon composition has already been formed, but their thermodynamic parameters (temperature T, pressure P) are the same for all nuclei-fragments and are determined by the characteristics of the initial or original nucleus. In this case, the task of finding the MCSFF is reduced to the analysis of the ordering of the canonical ensemble of constant pressure, formed by the totality of all possible two- or three fission fragments of the initial nucleus. The last one may be considered as the thermostat. On this stage, the theory allows one to expand the size of the ensemble, by including nuclei-fragments, are obtained as the results of fission neutrons and beta-particles emission with different length of chains. Moreover, one can regulate the range of these nuclear particles chains, excluding short-lived isotopes from the ensemble of fission's fragments and investigate the change of the MCSFF topology for different time intervals after the fission of the initial nucleus.

In the case of two fragments' scheme of the nucleus fission, for the i-th cluster, contains $N_{p,i}^{(j)}$ protons, $N_{n,i}^{(j)}$ neutrons in the j-th fragments, j = 1, 2 that are formed after considering emission the $n_i$ – beta particles ($^{\pm}\beta$) and $m_i$ –neutrons fission should satisfy the following conservation conditions:

$$\sum_{j=1,2} N_{p,i}^{(j)} + N_{n,i}^{(j)} + m_i = A_0,$$
$$\sum_{j=1,2} N_{p,i}^{(j)} + n_i = Z_0, \qquad (1)$$

where the positive sign of $n_i$ has the meaning the number of emitted $^{+}\beta$ and negative, respectively, the number of $^{-}\beta$ particles. Further calculation is based on the following assumptions:
- Characteristics of MCSFF are determined from the condition of thermodynamics ordering of the canonical ensemble which contains all nuclei-fragments clusters;
- Nucleons with different binding energy must be considered as statistically non-equivalents when calculating the configuration entropy, S, of the nuclear clusters;
- The nuclear particles of emission (neutrons of fission, beta-particles, gamma radiation) don't influence the constancy of the thermodynamic parameters of the nuclei fragments ensemble, namely, system's thermostat;

We account that the emission of fission's neutrons changes the volume V of the original nucleus, which leads to work done at a constant pressure of the nucleons in the nucleus [17]. To investigate the equilibrium condition of the ensemble of fission fragments, one must use the Gibbs thermodynamic potential [18]:

$$G = U - TS + P\Delta V, \qquad (2)$$

where $P\Delta V$, - work under constant pressure. The U values determined by the binding energy of the two-fragment cluster, its discrete spectrum $\{\varepsilon_i(V)\}$ is an additive quantity in the binding energy of the j -th nucleus fragment from the i -th cluster and have the negative values corresponding to the bound states of the nucleons:

$$\varepsilon_i = \sum_{j=1,2} \cdot \sum_{\langle N_p \rangle_i} \cdot \sum_{\langle N_n \rangle_i} U_j(A_{j,i}, Z_{j,i}) \qquad (3)$$

the symbol <...> means that summation in (3) is taken over the numbers of protons/neutrons, $N_{j,i}^p / N_{j,i}^n$, satisfying the conservation conditions (1). The configurational entropy S in (2)

$$S = \ln(w_i), \qquad (4)$$

calculated through the degeneracy factor $\omega_i$ that takes into account the statistical nonequivalence of nucleons with different specific binding energy in the fission fragments.

$$w_i = A_0! / (\prod_{j=1,2} (N_p^{(j)}! N_n^{(j)}! K(n_{i,j}))), \qquad (5)$$

where $K(n_{i,j})=1/n_{i,j}!$, $\prod_{j=1,2} x_j! = x_1! x_2!$. From the (5), one can see that entropy term in eq. (2) reaches maximum if $N_p^j = N_n^j$ and is responsible for the symmetrization of the fission yields with increasing of the nuclear temperature T.

For the matching of thermodynamic quantities and statistical averages, it is necessary to write the probability of realization, for example, of the two fragments cluster through the isobaric distribution function [18]:

$$f_i(V) = \omega_i \exp\{-(\varepsilon_i + P\Delta V)/T\}/Z_p, \quad (6)$$

where $\varepsilon_i$ - is one of the spectrum values of the initial energy (3) of the two fragment cluster, $w_i$ - determines its degeneration by the formula (5), and the statistical sum of $Z_p$ is calculated as:

$$Z_p = \sum_{k,V} \omega_k \exp\{-(\varepsilon_k + P\Delta V)/T\}.$$

The next step is the going from the probability of the i-th nuclear cluster formation to the distribution function $F(A_i)/F(Z_i)$ or the probability of the yield of a single fission's fragment with the mass ($A_i$)/charges ($Z_i$). This procedure can be done as follows: from (6), all initial (non-normalized) values of $F(A_1)$ are determined as the sum of probabilities of nuclear clusters, each of which contains a fragments fission of mass $A_1$, but possibly with different $Z_1$. A similar procedure is valid for establishing $F(Z_1)$. The next step is to apply the normalization procedure and determine the final and normalized values of $F(A_1)$ and $F(Z_1)$. These functions must satisfy the following equations:

$$\sum_{<A_1>} F(A_1) = \sum_{<Z_1>} F(Z_1) = 200\%,$$

where the symbols $<A_1>$, $<Z_1>$ have the same meaning as the conservation conditions in equation (3). Since the proposed method is based on thermodynamic principles, it does not contain time variables and not consider the kinetics of the fission process. However, this can be done indirectly, going in the calculation (6) from an ensemble containing all possible post-scission nuclear fragments (long and short-lived) to an ensemble of nuclear clusters containing only long-lived fission fragments. The latter case simulates the conditions for post-fission experiments when measurements are carried out after a long "cooling" time of the irradiated samples [19].

This method can take into account other important details post-scission process as the deformation of the fission fragments and the role of filled nuclear shells or magic numbers in the formation of the fission's channels indirectly, through mass formulas tabulated, for example, in recent reviews [20, 21].

### 3. Results and discussion

As an example, the possibilities of the method have been investigated by the example of the even proton, even neutron isotope $^{232}$Th, whose MCSFF has been well established [19]. As mentioned above, the taking into account the emission of elementary particles was carried out by including in the post-scission ensemble of fission fragments the nuclei formed as a result of the beta decay with chains by capacity (n), that is, in the interval [0, n] for $^+\beta$ and [-n, 0] for $^-\beta$ particles, and also fission neutrons with a capacity (m), or for intervals [0, m]. The calculation carried out for the $^{232}$Th isotope, shows that as a result of the emission of the nuclear particles both the beta particles ($^-\beta$) and the fission neutrons and enlarging of the fission fragments ensemble may change the nucleon composition of the most probable pairs of fragments of fission. Thus, as a result of the emission of nuclear particles, both the beta particles ($^-\beta$) and the fission neutrons in the ensemble increase the number of nuclides with nuclear shells close to the "magic" numbers, which have a greater specific binding energy in the spectrum of values $\{\varepsilon_i\}$. The role of neutrons emission is to perform isobaric work with a decrease in the volume of nuclei and leads to increase the configuration entropy (5) of the nuclear fission cluster. Formally, as can be seen from (6), the most probable are two fragments fission clusters that realize the minimum of the thermodynamic potentials of Gibbs (1) for a constant pressure ensemble. The results of such calculations are given in Tables 1 and 2, which contain 5 of the most probable two fragments clusters, both without, and when taking into account the emission of nuclear particles after the fission of 232Th. The calculation is given in the "cold" fission assumption with the value of the nuclear temperature T = 0.5 MeV. The sign near n indicates the sort of the emitted beta particles: -n suggests that the chains of -β particles in the range [n, 0] are taken into account; positive values relate to + β particles and their emission intervals [0, n].

Table 1 Evolution of the five most probable two fragment clusters for different length of chains of fission's neutrons: m defines all their sets from the intervals [0, m]. There is no beta emission.

| N\m | 0 | +1 | +2 | +3 | +4 |
|---|---|---|---|---|---|
| 1 | $\{^{100}_{40}Zr, ^{132}_{50}Sn\}$ | $\{^{100}_{40}Zr, ^{130}_{50}Sn\}$ | $\{^{96}_{40}Zr, ^{132}_{50}Sn\}$ | $\{^{96}_{40}Zr, ^{130}_{50}Sn\}$ | $\{^{96}_{40}Zr, ^{130}_{50}Sn\}$ |
| 2 | $\{^{102}_{40}Zr, ^{130}_{50}Sn\}$ | $\{^{98}_{40}Zr, ^{132}_{50}Sn\}$ | $\{^{94}_{38}Sr, ^{134}_{52}Te\}$ | $\{^{96}_{40}Zr, ^{132}_{50}Sn\}$ | $\{^{102}_{40}Zr, ^{132}_{50}Sn\}$ |
| 3 | $\{^{101}_{40}Zr, ^{131}_{50}Sn\}$ | $\{^{100}_{40}Zr, ^{132}_{50}Sn\}$ | $\{^{98}_{40}Zr, ^{130}_{50}Sn\}$ | $\{^{94}_{38}Sr, ^{134}_{52}Te\}$ | $\{^{94}_{38}Sr, ^{134}_{52}Te\}$ |
| 4 | $\{^{104}_{40}Zr, ^{128}_{50}Sn\}$ | $\{^{101}_{40}Zr, ^{130}_{50}Sn\}$ | $\{^{100}_{40}Zr, ^{130}_{50}Sn\}$ | $\{^{104}_{40}Zr, ^{128}_{50}Sn\}$ | $\{^{98}_{40}Zr, ^{130}_{50}Sn\}$ |
| 5 | $\{^{99}_{39}Y, ^{133}_{51}Sb\}$ | $\{^{101}_{40}Zr, ^{130}_{50}Sn\}$ | $\{^{97}_{40}Zr, ^{132}_{50}Sn\}$ | $\{^{98}_{40}Zr, ^{130}_{50}Sn\}$ | $\{^{97}_{40}Zr, ^{130}_{50}Sn\}$ |

Table 2 The evolution of the five most probable two fragment clusters for different length of beta-decay chains: n defines their set from the intervals [0, n], -n was used for the chains of -β particles in the range [n, 0]. There are two cases where the emission of fission neutrons is absent (m = 0), and for a chain of fission neutrons with capacity (m = 3), that is, for their intervals [0, 3].

| | N\n | -4 | -3 | -2 | -1 | 0,+1,+2,+3,+4 |
|---|---|---|---|---|---|---|
| | 1 | $\{^{106}_{44}Ru, ^{126}_{50}Sn\}$ | $\{^{105}_{43}Tc, ^{127}_{50}Sn\}$ | $\{^{104}_{42}Zr, ^{128}_{50}Sn\}$ | $\{^{102}_{41}Nb, ^{130}_{50}Sn\}$ | $\{^{100}_{40}Zr, ^{132}_{50}Sn\}$ |
| | 2 | $\{^{108}_{44}Ru, ^{124}_{50}Sn\}$ | $\{^{104}_{43}Tc, ^{128}_{50}Sn\}$ | $\{^{102}_{42}Mo, ^{130}_{50}Sn\}$ | $\{^{101}_{41}Nb, ^{131}_{50}Sn\}$ | $\{^{102}_{40}Zr, ^{130}_{50}Sn\}$ |
| m=0 | 3 | $\{^{104}_{44}Ru, ^{128}_{50}Sn\}$ | $\{^{101}_{42}Zr, ^{131}_{51}Sb\}$ | $\{^{98}_{40}Zr, ^{134}_{52}Te\}$ | $\{^{100}_{40}Zr, ^{132}_{51}Sb\}$ | $\{^{101}_{40}Zr, ^{131}_{50}Sn\}$ |
| | 4 | $\{^{96}_{40}Zr, ^{136}_{54}Xe\}$ | $\{^{100}_{42}Zr, ^{132}_{51}Sb\}$ | $\{^{106}_{42}Mo, ^{126}_{50}Sn\}$ | $\{^{100}_{41}Nb, ^{132}_{50}Sn\}$ | $\{^{104}_{40}Zr, ^{128}_{50}Sn\}$ |
| | 5 | $\{^{114}_{46}Pd, ^{118}_{48}Cd\}$ | $\{^{106}_{42}Tc, ^{126}_{50}Sn\}$ | $\{^{100}_{42}Mo, ^{132}_{50}Sn\}$ | $\{^{99}_{40}Zr, ^{133}_{51}Sb\}$ | $\{^{99}_{39}Y, ^{133}_{51}Sb\}$ |

The most probable nuclear clusters are given in Tables 1, and 2 were obtained at the condition of the minimum of the thermodynamic potential of Gibbs (2), whose parameters were changes to account the different chains of emitted nuclear particles. The ensemble of fission fragments is supplemented by most probable clusters containing more stable isotopes. It is interesting to investigate the influence of how the emission of elementary particles causes the changes ensemble of post-scission nuclear clusters. These changes are followed from the potential G. Since G is in the exponent of the isobaric distribution function (5) its smallest changes can significantly change of content and ratio of the probable two-fragments fission clusters.

Figure 1 shows the dependence of the values of the Gibbs potential G (2), normalized to A0 as a function of length of the chains of emitted beta-particles n for the five most probable two fragments fission clusters from the Tables 1, 2. As well as the results in Table 1, 2, the dependences of potentials G are obtained without (m = 0), and with the account of the neutron fission chain with a capacity of m = 3. The inset shows the dependence of G when taking into account the chains of fission neutrons with a capacity of m from 0 to 3. The numbers of the curves correspond to the values of N from tables 1.2. As can be seen, the emission of electrons ($^-\beta$) in comparison with the same volume of emission $^+\beta$ particles or fission neutrons leads to more significant changes of function $G$ and the realization more stable states of the ensemble of nuclear clusters. Note that such changes do not exceed 1% for the G module; however, as mentioned above, the values of the exponent of expression (5) are very sensitive to its changes. Therefore, even minor changes in its magnitude can significantly affect the character of the thermodynamic ordering and the redistribution of probable yields in the MCSFF. As can be seen, the positron emission ($^+\beta$) almost does not lead to changes G and redistribution of MCSFF, and the emission of neutrons causes an effect on G less than four times that of $^-\beta$ particles emissions.

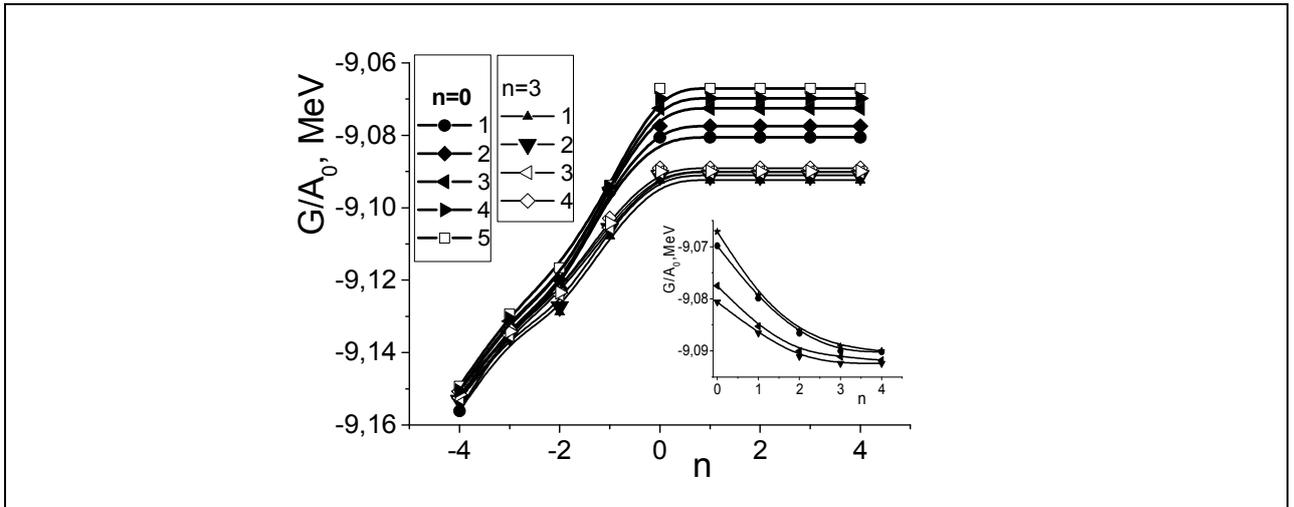

Fig.1 The dependencies of the reduced potential of Gibbs for the most probable nuclear clusters from Tabl.1, on the length of the chains of emitted beta-particles n. Dependences were obtained without (m = 0), and when considered the neutrons fission with the chain's capacity m = 3. The inset shows the same dependence for G taking into account chains of neutron-fission with the chain's capacity of m from 0 to 3.

The emission both the fission's neutron (m = 3), and the beta-particles ($^-\beta$) leads to reducing the changes by almost 75% of G modulus compared to the case where the fission neutrons are not taken into account. The analysis shows that consideration the emission of nuclear particles also leads to domination in the spectra of the fission fragments of more stable nuclei with their half-life of several seconds at n = 0, m = 0 to 371.8 days and 2.30 × 10$^5$ years as $\{^{106}_{44}Ru, ^{126}_{50}Sn\}$ for the chain of particles with a capacity of 4, m = 0, Tab. 2. The same tendency occurs when one account the emission of neutron fission only.

As can be seen from Table 1 and 2, taking into account only neutron emission retains the position of the peak of heavy fission fragments, especially for the charge spectra formed mainly by the isotopes $^{130,132,134}$Sn, and only for m = 3, 4 in the spectra had been appearing isotopes $^{134}$Te. The peak of light fission fragments are mainly formed by $^{94-104}$Zr isotopes and are very sensitive to the post-fission emission of beta particles. In this case, one can see that Zr isotopes are replaced by $^{100-102}$Nb, $^{100,102,106}$Mo, $^{104,106,108}$Ru, as well as the isotopes Tc and Pa. Moreover, the inclusion of longer chains of emission of nuclear particles in the fission leads to the symmetrization of the MCSFF's topology. This fact is especially evident in the example of the charge spectra and the emission of beta particles ($^-\beta$).

The redistribution of yields of the most probable fission fragments, as shown in tables 1 and 2 leads to a change in the topology of the MCSFF. Figure 2 shows the results of calculations MCSFF for the above cases. First of all, it can be stated that the emission of $^+\beta$ particles does not lead to changes in the yields of the MCSFF, but taking into account the presence of post – fission $^-\beta$ particles substantially change their topology. In other words, there are the tendencies towards the domination the symmetric fission mode over the asymmetric one when the two-hump topology of the MCSFF goes into one-hump, and this trend is enhanced in case the account both the emission of beta-particles and neutron fission. As can be seen from the Fig.2, such a symmetrization has a nonmonotonic behavior with increasing the chains' lengths of emitted beta-particles, n. It has its features for intervals of emitted $^-\beta$ particles [-3, -2], when in the groups of the most probable nuclear clusters, the peak of heavy fission fragments starts to be formed not only with isotopes Sn with filled charge shells (Z = 50) but also with isotopes $^{132,134}$Te, $^{131,132}$Sb, $^{134,135}$I. The peak of probable light fission fragments, which at n = 0 was formed by isotopes Zr when considering the emission of nuclear particles, are created by the isotopes Mo, Tc, Pa and Ru with their tendency to the filled neutron shells N = 50. When one considers the neutron fission chains only, the no monotonic behavior of the MCSFF also occurs for m interval [2,3]. From Table 1, one can see that in this case, the peak of the light fission fragments changes their probable the Zr isotope on Sr ones. In the peak of the heavy fission fragments yields, instead of the isotopes of Sn, the mentioned isotopes Te appear.

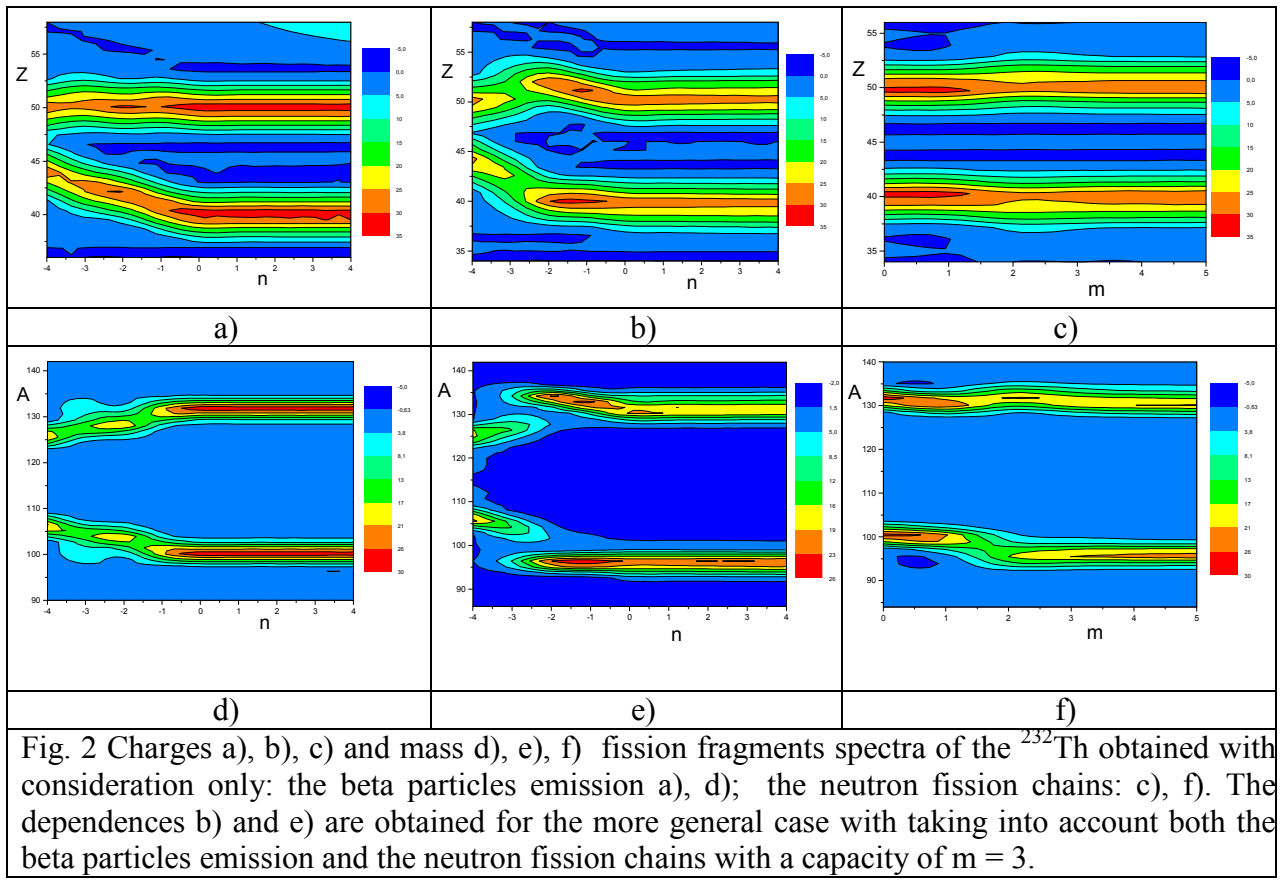

Fig. 2 Charges a), b), c) and mass d), e), f) fission fragments spectra of the $^{232}$Th obtained with consideration only: the beta particles emission a), d); the neutron fission chains: c), f). The dependences b) and e) are obtained for the more general case with taking into account both the beta particles emission and the neutron fission chains with a capacity of m = 3.

As indicated above, entropy factors are also affected on the MCSFF symmetrization, for example, due to the increase in the configuration entropy of the nuclear clusters, which reaches its maximum in the case $N_{p,i}^{(j)} \cong N_{n,i}^{(j)}$ for the j-th fragment, see (3). Therefore, the results of calculations of the fragments fission yields for 232Th presented in Fig. 2, to enable experimenters to estimate the possibilities of their methods for study the MCSFF. Analysis of Fig. 2 shows that considering the emission of nuclear particles under the fission of the atomic nuclei leads to a disturbance of the symmetry of the MESF's yields, when, for example, the proportion of the number of 2 nuclear clusters forming the peak of the light and heavy nuclei-fragments are different. For example, for charge spectra, the channel of heavy fission's fragments for a magic number Z = 50 is formed by the biggest number of two fission fragment clusters than the corresponding light (Z = 43-41). For mass spectra, the channel A = 132 is dominant, being formed by bigger chains of nuclei-fragments are closer to twice the magic number isotope Sn (), than the corresponding genetically connected channels A = 96-100, which the observable when one takes into account the fission's neutrons.

And, finally, there are the problems of comparing the calculated and experimentally observed the MCSFF of atomic nuclei, Fig. 2. To this end, it is necessary always to take into account the abilities of experimental methods for this sort of investigations, for example, that concerning $^{232}$Th, see [19].

First of all, this is due to the experimental possibility for nuclear fission studies, either immediately after the scission experiments, or after some time when the ensemble the two-fragments' clusters have expended on the nuclei formed as a result of emission of chains of the nuclear particles. Figure 2 shows that in these cases, the MCSFF is formed as a result of the competition of symmetric and asymmetric fission modes, and, over time, the dominance of the symmetrical mode over the asymmetric one is dominant. On the other hand, the gamma-spectrometric method is valid only on the totality of gamma-active nuclei all of which are the fission fragments and the mass-spectrometric ones would give their error in the presence of ultra-short living nuclides, the half-life of which is determined by the emission of charged parts, or neutrons.

Therefore, the results of calculations of the fragments fission yields for $^{232}$Th presented in Fig. 2, enable experimenters to estimate the possibilities of their methods for study the MCSFF.

## 4. Conclusions

Thus, the results of the study of the transformation of the MCSFF carried out on the example of the $^{232}$Th fission's fragments show the ability to investigate the nature of the change in its topology by taking into account the emission of post – scission's nuclear particles. Although the proposed statistical approach, are aimed at studying the thermodynamic ordering of the ensemble of fission's fragments and does not contain time variables, the method proposed allows one indirectly investigate the time evolution and the nature of their transformations.

The proposed statistical method allows also us to find the most probable two fragments' nuclear clusters formed in the fission of atomic nuclei and to investigate their in time evolution through taking account the different length of elementary particles as beta and neutrons ones had been emitted after the scission of the initial nuclei. It is also obtained that the post scission nuclear particles emission leads to the stimulation of the symmetric fission mode above the asymmetric one when forming the MCSFF.

These results are also important to testify the experimental abilities for the MCSFF investigation, in particular, by taking into account the "cooling" time in post-fission experiments, the identification of gamma-active nuclides, or the possibilities of mass spectrometric techniques.

The authors are grateful V. Denysov for assistance and discussions with this work.